\documentclass[sigconf]{acmart} 

\renewcommand\footnotetextcopyrightpermission[1]{} 
\AtBeginDocument{%
	\providecommand\BibTeX{{%
			\normalfont B\kern-0.5em{\scshape i\kern-0.25em b}\kern-0.8em\TeX}}}

\usepackage[utf8]{inputenc} 
\usepackage[T1]{fontenc}    
\usepackage{hyperref}       
\usepackage{url}            
\usepackage{booktabs}       
\usepackage{amsfonts}       
\usepackage{nicefrac}       
\usepackage{microtype}      

\usepackage{booktabs} 
\usepackage{bbm}
\usepackage{amsmath}
\usepackage{amsthm}
\usepackage{enumitem}
\usepackage{array}
\usepackage{graphicx}
\usepackage[linesnumbered,lined,boxed,commentsnumbered,ruled,vlined]{algorithm2e}
\usepackage{subfigure}
\usepackage{booktabs} 
\usepackage{mathtools} %
\usepackage{balance}
\usepackage{multirow}

\usepackage{mathtools}
\DeclarePairedDelimiter\ceil{\lceil}{\rceil}

\begin{document}
	
	\settopmatter{printacmref=false} 
	\fancyhead{}
	
	

	\title{Learning to Detect Few-Shot-Few-Clue Misinformation}
	
	\author{Qiang Zhang}
	\affiliation{%
		\institution{University College London}
		\city{London}
		\country{United Kingdom}
	}
\email{qiang.zhang.16@ucl.ac.uk}
	
	\author{Hongbin Huang}
	\affiliation{%
		\institution{Sun Yat-sen University}
		\city{Guangzhou}
		\country{China}
	}
	\email{huanghb27@mail2.sysu.edu.cn}

	\author{Shangsong Liang}
	\affiliation{
		\institution{Sun Yat-sen University}
		\city{Guangzhou}
		\country{China}
	}
	\email{liangshangsong@gmail.com}
	
		\author{Zaiqiao Meng}
	\affiliation{%
		\institution{University of Cambridge}
		\city{Cambridge}
		\country{UK}
	}
	\email{huanghb27@mail2.sysu.edu.cn}

	\author{Emine Yilmaz}
	\affiliation{%
		\institution{University College London \& Amazon}
		\city{London}
		\country{United Kingdom}}
	\email{emine.yilmaz@ucl.ac.uk}
	
	\vspace{1em}

	\renewcommand{\shortauthors}{}
	
	\begin{abstract}
		A large volume of false textual information has been disseminating for a long time since the prevalence of social media. 
		The potential negative influence of misinformation on the public is a growing concern.
		Therefore, it is strongly motivated to detect online misinformation as early as possible. 
		\emph{Few-shot-few-clue learning} applies in this misinformation detection task when the number of annotated statements is quite few (called \emph{few shots}) and the corresponding evidence is also quite limited in each shot (called \emph{few clues}). 
		Within the few-shot-few-clue framework, we propose a {task-aware} Bayesian meta-learning algorithm to extract shared patterns among different topics ({i.e. \emph{different tasks}}) of misinformation. 
		Moreover, we derive a scalable method, i.e., amortized variational inference, to optimize the Bayesian meta-learning algorithm. Empirical results on three benchmark datasets demonstrate the superiority of our algorithm. This work focuses more on optimizing parameters than deigning detection models, and will generate fresh insights into data efficient detection of online misinformation at early stages.
		
		
	\end{abstract}


	\maketitle

\section{Introduction}

Online misinformation is affecting a wider range of individuals as an increasing number of people tend to use digital sources to access textual news. However, unlike the traditional news published by credible institutions, digital information on the web has not proven its quality. Individuals and organizations publish and share news on popular social media platforms, such as Facebook and Twitter, without accurate content checking. 
Due to financial or political purposes, some social media accounts publish biased and even false information around popular topics. 
Some other social accounts spread such misinformation to a wider range, intentionally (for those who know the news is fake) or unintentionally (for those who believe the news content is true). 

Online misinformation can be defined as made-up statements on the web that are verifiably false, and could mislead and deceive readers~\cite{allcott2017social}. 
%
%
Misinformation, when done on a large scale, can influence the public by depicting a false picture of the reality. 
Hence, detecting misinformation effectively has become one of the most severe challenges faced by social media platforms.

The task of misinformation detection aims to identify whether a statement is \emph{True} or \emph{False}. 
Researchers have been rectifying this epidemic in two ways.
One way is manual-checking that  has been tackled by some news websites, such as Snopes and FactCheck.
However, such {a} manual way is expensive 
to be able to check all the daily generated statements appearing on the web. 
Automatic tools, especially machine learning models, have been developed to accelerate the verification procedure~\cite{potthast2017stylometric}. 
Nonetheless, these models are usually data hungry and require considerable annotations. How to alleviate this requirement is of high research interest. 
Also, to mitigate negative influence of misinformation, detection should be conducted as early as possible. This puts forward another requirement on the detection models because clues, such as social engagements, are relatively rare at early stages. We refer to, the task of learning to detect misinformation meeting these two requirements, as the \emph{few-shot-few-clue} learning task, which is the main focus of this paper. 

\begin{figure*}[!t]
	\centering
	\vspace{-1em}
	\includegraphics[width=0.7\textwidth]{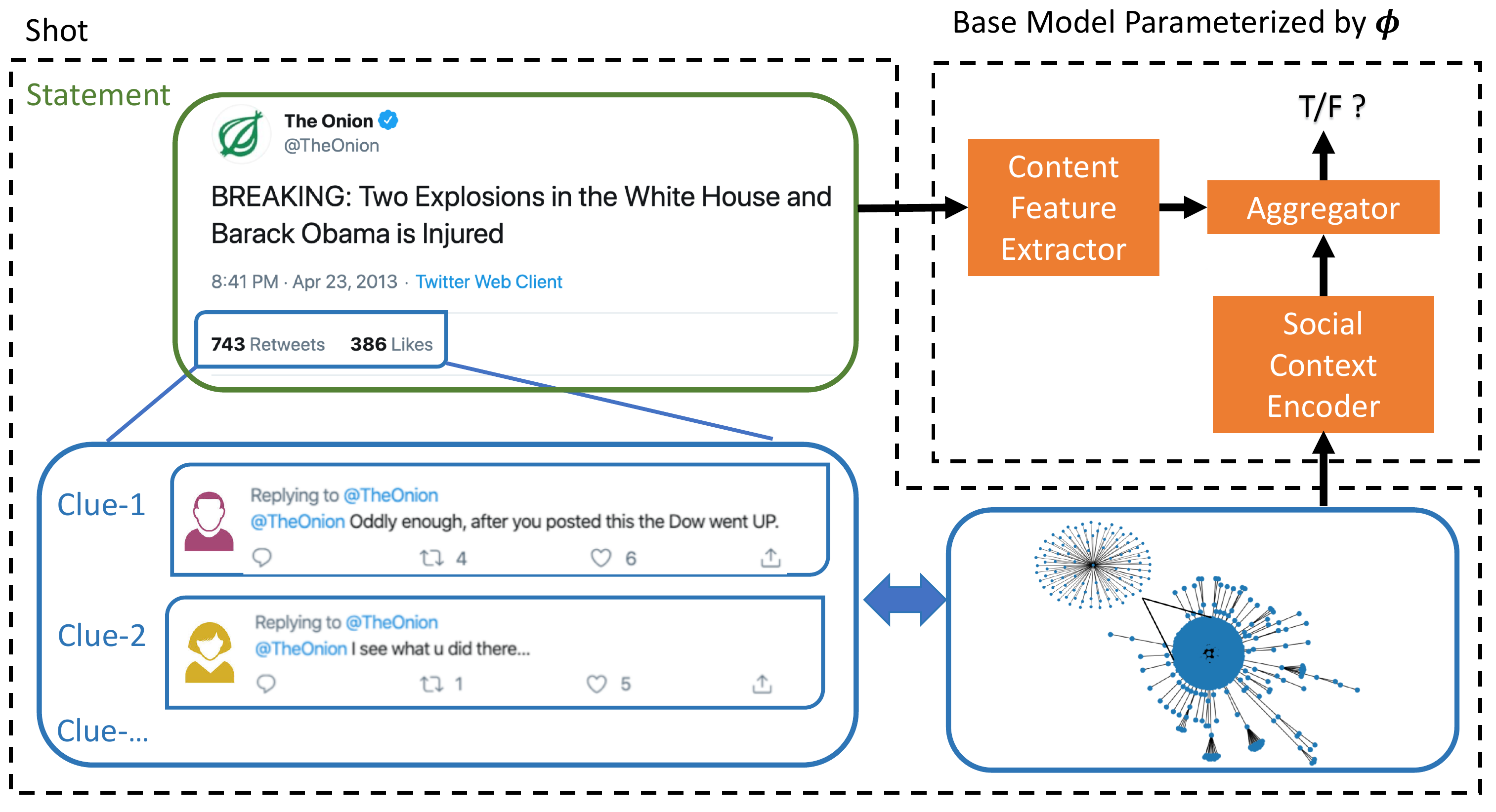}
	\caption{We show a shot, including a statement and accompanied clues with the equivalent propagation graph. The shot is input to a base model containing a content feature extractor, a social context encoder and an aggregator to identify veracity.}
	\label{fig:false_infor}
\end{figure*}

Previous works tried to either generally increase data efficiency of machine learning models~\cite{weiss2016survey}
or detect misinformation with limited clues at early stages~\cite{wang2018eann}. 
Notwithstanding, there has not been a unified framework for data efficient detection of misinformation at early stages. 
We systematically review previous works and propose a new learning framework for misinformation detection, i.e., \emph{few-shot-few-clue} learning.
This framework extends the conventional few-shot learning by incorporating the concept of \emph{clue}. As a shot (e.g., the statement and its social engagements in Figure~\ref{fig:false_infor}) is a datum, clue can be the evidential signal (e.g., social engagement behavior and commenting words in Figure~\ref{fig:false_infor}) inside the shot.
Few-shot-few-clue learning applies when there are a limited number of annotated shots and a limited number of clues in each shot.

Within the framework of few-shot-few-clue learning, we first define a topic-centered misinformation detection as a \emph{task}. Different tasks indicate misinformation detection of different events/topics (e.g., presidential election and COVID-19). 
{
It has been found that tasks share transferable patterns in misinformation detection, such as linguistic features of statements and stance patterns of engaged users in social platforms~\cite{potthast2017stylometric,wang2018eann}. 
}
Based on these findings, we propose a {task-aware} Bayesian meta-learning algorithm to recognize the shared patterns between various tasks and improve detection performance on new topics. Finally, for sake of scalable optimization of the meta-learning algorithm, we derive an amortized variational inference method.
To sum up, this study advances the state-of-the-art in four aspects:
\begin{enumerate}
	\item We define a new framework of few-shot-few-clue learning for misinformation detection at early stages. 
	\item We develop a {task-aware} Bayesian meta-learning algorithm to tackle the task of few-shot-few-clue misinformation detection.
	\item We derive an amortized variational inference method for scalable optimization of the meta-learning algorithm.
	\item We conduct a systematic experimentation on three real-world datasets. Empirical results demonstrate that our algorithm outperforms the state-of-the-art detection methods. 
\end{enumerate}

\noindent The remainder of the paper is organized as follows: \S~\ref{sec:related} summarizes the related work; 
\S~\ref{sec:preliminary} presents preliminary concepts of few-shot learning and meta-learning;
\S~\ref{sec:task} defines the new few-shot-few-clue learning framework; 
\S~\ref{sec:method} details the base model for misinformation detection and the proposed task-aware Bayesian meta-learning algorithm; \S~\ref{sec:optimization} derives the amortized variational inference method;
\S~\ref{sec:experiments} describes the used datasets and experimental setup; \S~\ref{sec:results} is devoted to experimental results; and \S~\ref{sec:conclusion} concludes the paper.

\section{Related Work}
\label{sec:related}
We provide a brief review of four lines of recent works on misinformation detection, early misinformation detection, transferable patterns of misinformation and meta-learning.

\subsection{Misinformation Detection Methods}
\label{subsec:MID}
%
\paragraph{\textbf{Content features}}
Misinformation is fabricated to mislead the public for financial or political gains, not to objectively report a fact. Hence they often contain opinionated or inflammatory language and images, which motivates content-based detection methods. In order to reveal linguistic differences between true and false statements, textual methods have studied the lexical, synthetic and topic features at the word, sentence and document levels~\cite{potthast2017stylometric}. 
%
Some sentiment features, such as positive words (e.g., love, sweet), negating words (e.g., not, never), cognitive action words (e.g., cause, know) and inferring action words (e.g., maybe, perhaps), are reported to help detect rumour~\cite{kwon2013prominent}. 
%
Sensational or even fake images provoke anger or other emotional response of consumers. For example, Deepfake~\cite{floridi2018artificial} employ deep learning methods to generate fake images and videos to convey misleading information.
Visual-based features are extracted from images and videos to capture the characteristics of misinformation.
Recently, various visual and statistical features have been extracted for news verification.
\citet{yang2018ti} developed a convolutional neural network to extract text and visual features simultaneously. 
~\citet{khattar2019mvae} propose a  multimodal variational auto-encoder to extract visual features and detect fake news.

\paragraph{\textbf{Social context}}
On social media platforms, every piece of news is correlated to other posts and users. Interactions among users and contents (e.g., commenting, reposting, tagging) provide rich reference evidence for misinformation detection. 
One type of interactions is post-based, which replies on users' commentary posts towards the original statement~\cite{zhang2018ranking,zhang2019stances}. 
People react to a piece of statement by expressing their stances or emotions in social media posts. 
Stances can be categorized as supportive, opposing, and neutral, which can be used to infer statement veracity.
As the attention  mechanism has led to improved performance in various areas~\cite{zhang2018variational,zhang2020sahp,zhang2021learning},
this observation motivated the development of an attention-based recurrent neural network (RNN) model.
~\citet{zhang2019reply} propose a probabilistic deep learning model to utilize user replies as auxiliary evidence.
~\citet{yang2019unsupervised} consider statement veracity and user credibility as latent variables to predict their stances towards statement veracity. %
Reinforcement learning has also been tried to provide weak supervision in~\citet{wang2020weak}.
The other interaction type is statement propagation over time.
~\citet{P18-1184} learns features of non-sequential propagation structure of tweets.
A top-down and a bottom-up recursive neural networks were used to predict statement veracity. 
\citet{monti2019fake} harness Graph Convolutional Networks (GCN) to encapsulate the propagation structure of heterogeneous data.
%
\citet{lu2020gcan} propose Graph-Aware Co-Attention Network (GCAN) while\citet{zhang2021detecting} uses temporal point processes for dynamic engagement modeling. 
\subsection{Early Misinformation Detection}
Misinformation is created to mislead the public and can lead to severe consequences if widely propagated. Therefore, early detection is crucial.
Convolutional neural networks (CNN) are used to extract interaction features between temporally sequential posts.
~\citet{nguyen2017early} propose another CNN-based model to learn latent representations of tweets and predict event veracity by aggregating predictions of all related statements.
~\citet{liu2018mining} find that only a part of comments on social platforms are helpful to classify whether a statement is fake. Then they 
design an attention-based detection model to evaluate relative importance of comments according to their attention values. 
They also conclude that the attention mechanism contributes to early misinformation detection. 
~\citet{liu2018early} use recurrent and convolutional networks to extract propagation features while ~\citet{shu2020leveraging} leverage weak social supervision from multiple sources for early detection.

\subsection{Transferable Patterns}
Misinformation detection models may not generalize well across topics and cultures~\cite{horne2020all}, therefore
transferable patterns have attracted researchers' interests.
~\citet{potthast2017stylometric} have analyzed the similarity of the writing style of hyperpartisan news and its connection to misinformation.
~\citet{castelo2019topic} consider the changing discourse of emerging events, and propose a topic-agnostic approach that uses linguistic and web-markup features to identify misinformation.
~\citet{huang2020conquering} point out that the word/topic distribution of articles from different media sources may be different, so they develop a synthetic network that focuses on function words and syntactic structures for a more generalized representation. 
~\citet{wang2018eann} propose an event adversarial neural network where an event discriminator component aims to remove event-specific features.
All these methods aim for features, while our work aims for model parameters that allow for strong generalization.
%
%
%

\subsection{Meta-Learning}
Fast learning is a hallmark of human intelligence. 
Due to high efficiency of recognizing shared patterns,
meta-learning has been drawing increasing research attention~\cite{lemke2015metalearning}. 
There are different approaches to meta-learning: metric learning-based, optimization-based and probabilistic.
The first approach to meta-learning is based on metric learning in the embedding space. ~\citet{vinyals2016matching}
propose to match the embeddings of query examples with those of the support ones. ~\citet{zhou2018deep} propose to learn a concept space where the concept matching of categories is conducted.
One optimization-based work in~\cite{finn2017model} is model-agnostic meta-learning (MAML) algorithms that learns a proper model initialization. Through a few steps of gradient descent-based fine-tuning, the initialization can be adapted to different tasks. From the Bayesian point of view, probabilistic meta-learning is studied in~\cite{ravi2018amortized},
where the initialization is treated as the prior of model parameter distributions. The posterior distributions are derived from the dataset of each learning task. 
We customize meta-learning by using topic similarity to adaptively control how the meta parameter will initialize the topic-specific parameter.
One concurrent work~\cite{wangmulti} combines MAML and neural processes to deal with misinformation. We differ from this work from two aspects: (1) 
the proposed {task-aware} Bayesian meta-learning algorithm can provide uncertainty estimation that is infeasible for optimization-based meta-learning including MAML, and is less prone to be under-fitting than neural processes; and (2) we use topic similarity when adapting global knowledge to tasks, which addresses the concern of dissimilar task distributions in~\cite{wangmulti}. 


\section{Preliminaries}
\label{sec:preliminary}

Supervised deep learning usually suffers from the desideratum of a large-scale of annotated dataset. However, labels for some tasks are usually limited.
To improve data efficiency and generality of neural networks, there have been research efforts on how to utilize the transferable patterns between different learning tasks. 
One typical view is to recognize such patterns from a set of tasks that enable efficient learning of new unknown tasks. 
The \emph{few-shot learning} problem specifies scenarios where there are limited annotated data per class in classification tasks.  
Few-shot classification is an instantiation of meta-learning under the paradigm of supervised learning.

In the regime of meta-learning, there are multiple learning tasks drawn from a certain distribution $p(\tau)$. 
The key assumption of meta-learning is that tasks from this distribution share common patterns.
The goal of meta-learning is to discover such patterns by training a model on multiple tasks from the distribution $p(\tau)$.
To this end, the model is trained during the so-called meta-training step on a meta-training set that includes multiple tasks, and is evaluated during the meta-test step.

Suppose {a} task $\tau$ is a supervised learning problem, and the training and test datasets of this task can be represented as $\mathcal{D}^{\tau}=\left\{\mathbf{X}^{\tau}, \mathbf{Y}^{\tau}\right\}$ and $
\tilde{\mathcal{D}}^{\tau}=\left\{\tilde{\mathbf{X}}^{\tau}, \tilde{\mathbf{Y}}^{\tau}\right\}$ respectively. 
Commonly, there are only a few labeled data points in the training set $\mathcal{D}^{\tau}$.
The meta-training set $\mathcal{D}_{\rm{meta-train}}$ is made up of multiple tasks from $p(\tau)$, i.e., $\mathcal{D}_{\rm{meta-train}}=\{(\mathcal{D}^{1},\tilde{\mathcal{D}}^{1}), (\mathcal{D}^{2},\tilde{\mathcal{D}}^{2}), \cdots, (\mathcal{D}^{\mathcal{T}},\tilde{\mathcal{D}}^{\mathcal{T}})\}$, where $\mathcal{T}$ denotes the number of tasks used during meta-training.
While during meta-test, one or multiple new tasks can be used; for simplicity and without loss of generality, we use one task to denote the meta-test set $\mathcal{D}_{\rm{meta-test}}=\{ (\mathcal{D}^{\mathcal{T}+1},\tilde{\mathcal{D}}^{\mathcal{T}+1})\}$.

The learning procedure of meta-learning algorithms is split into two steps. First, meta-learning aims to learn the meta-parameter $
\boldsymbol{\theta}$ from $\mathcal{D}_{\rm{meta-train}}$, 
\begin{equation}
\boldsymbol{\theta}^{\star}  = \mathop{\arg\max}_{\boldsymbol{\theta}} \log p(\mathcal{D}_{\rm{meta-train}} \mid \boldsymbol{\theta}).
\end{equation}
Using $\boldsymbol{\theta}^{\star}$ as prior, the second step is to obtain the task-specific model parameter $
\boldsymbol{\phi}$ that is able to generalize well on $\mathcal{D}_{\rm{meta-test}}$ after a few trials,
\begin{equation}
\boldsymbol{\phi}^{\star}  = \mathop{\arg\max}_{\boldsymbol{\phi}} \log p( \mathcal{D}_{\rm{meta-test}} | \boldsymbol{\phi},  \boldsymbol{\theta}^{\star}).
\end{equation}

The meta-parameter $\boldsymbol{\theta}^{\star} $ can be viewed as the shared feature patterns among tasks from $p(\tau)$. After slight modifications, $\boldsymbol{\theta}^{\star}$ builds the internal representation $
\boldsymbol{\phi}^{\star} $ that is suitable for new tasks and produces promising results.
Meta-learning allows quick model adaptation from a small number of training data, and continuous adaption as more data becomes available. 
It is fast and flexible learning and avoid overfitting to new information.

\section{Problem formulation}
\label{sec:task}
In this section, we first connect few-shot learning to misinformation detection, then extend few-shot learning to a new \emph{few-shot-few-clue} learning framework that is formulated for efficient detection of misinformation at early stages. 
Table~\ref{tab:symbol} shows the symbols used in this paper.

\begin{table}[t]
	\centering
	\caption{Definitions of symbols that are used in the paper.}
	\vspace{-1em}
	\label{tab:symbol}
	\begin{tabular}{ll}
		\toprule
		Symbol & Description \\ 
		\hline
		$\tau$ & a task of topic-centered misinformation detection \\
		$p(\tau)$ & the distribution of tasks \\
		$\mathcal{T}$ & the total number tasks \\
		$m$ & a shot consisting of a statement and clues \\
		$M^\tau$ & the total number of shots of the task $\tau$ \\
		$\boldsymbol{s}_m^\tau$ & the statement in the shot $m$ of the task $\tau$\\
		$S^\tau$ & the set of statements of all the shots of the task $\tau$ \\
		$y_m^\tau$ & the label in the shot $m$ of the task $\tau$ \\
		$Y^\tau$ & the set of labels of all the shots of the task $\tau$ \\
		$\boldsymbol{c}_{m}^\tau$ & the set of clues in the shot $m$ of the task $\tau$ \\
		$\boldsymbol{c}_{m,n}^\tau$ & the $n$-th clue in the shot $m$ of the task $\tau$ \\
		$N_m^\tau$ & the total number of clues in the shot $m$ of the task $\tau$ \\
		$C^\tau$ & the set of clues of all the shots of the task $\tau$ \\
		$\mathcal{D}^{\tau}$ & the training dataset for the task $\tau$\\
		$\tilde{\mathcal{D}}^{\tau}$ & the test dataset for the task $\tau$\\
		$\boldsymbol{\phi}$ & the local model parameter for the task $\tau$\\
		$\boldsymbol{\theta}$ & the global model parameter shared across all tasks\\
		\bottomrule
		\vspace{-1em}
	\end{tabular}
\end{table}

\subsection{Few-Shot Misinformation Detection}
Annotating statement veracity is expensive and time-consuming. 
This leads to the challenge of how to improve efficiency of annotated data.
Section~\ref{sec:preliminary} discusses {the few-shot classification task} for a few labeled examples per category, the key of which is to properly recognize transferable patterns among different tasks, e.g., linguistic features and user stances~\cite{potthast2017stylometric,wang2018eann}.
Thus, the challenge of data efficiency can be tackled by few-shot learning.
From this viewpoint, we formulate misinformation detection as a few-shot learning problem. 
As usual, the concept ``few'' indicates a specific number such as 5.
For a statement $\boldsymbol{s}$ and clues $\boldsymbol{c}$, the veracity $y$ is predicted by a model parameterized by $\boldsymbol{\phi}$. The model parameter $\boldsymbol{\phi}$ for different tasks shares the same prior $\boldsymbol{\theta}$. To improve data efficiency and convenient adaptation to unseen events/topics, misinformation detection models need to extract salient patterns, i.e., $\boldsymbol{\theta}$, that can be transferred across topics and events.


\subsection{Few-Shot-Few-Clue Learning}
\begin{figure}[t]
	\center
	\includegraphics[width=0.7\columnwidth]{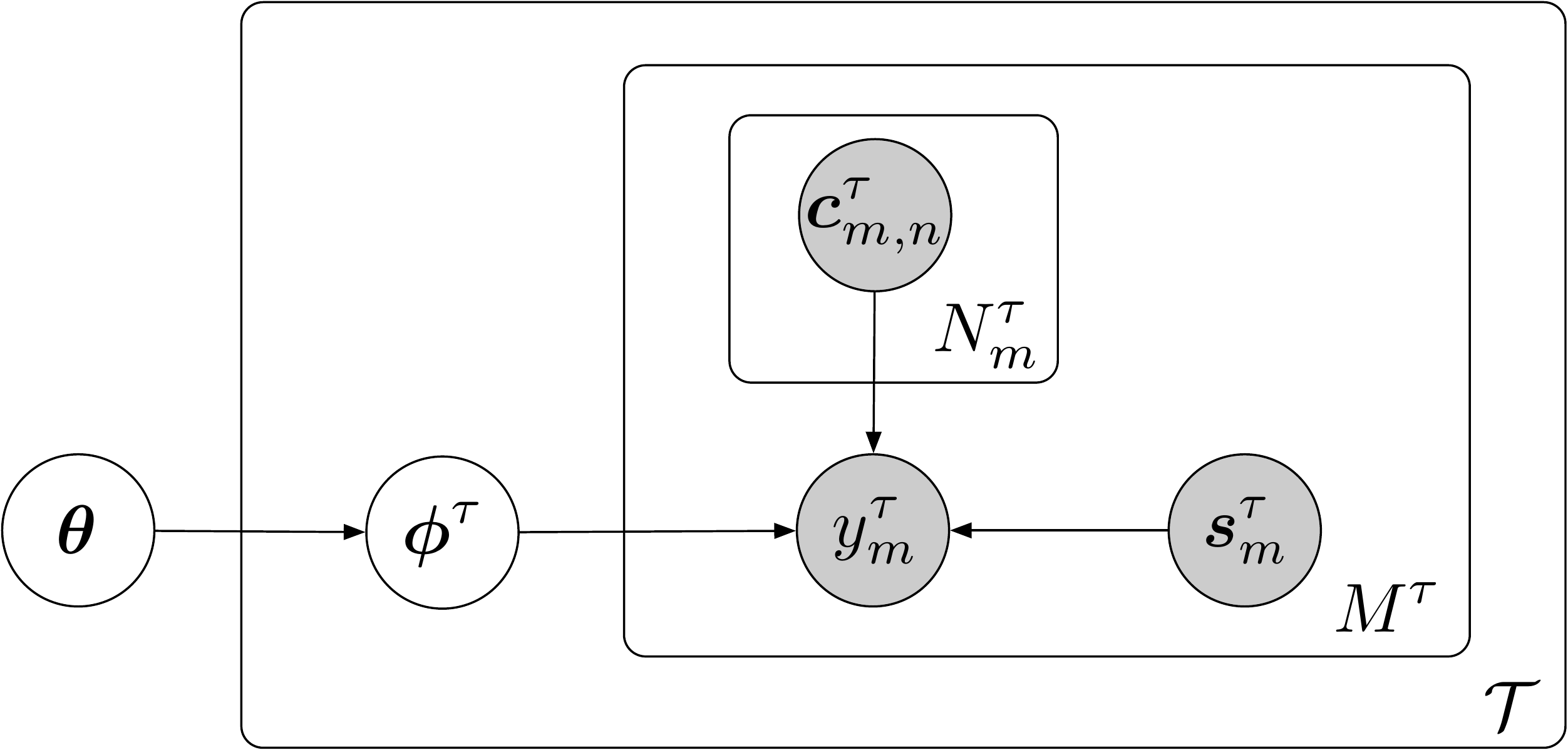}
	\caption{The directed graphical model. Grey and white nodes represent observed and latent variables, respectively. 
		The variable $\boldsymbol{c}_{m}^\tau$ is decomposed into a set of clues $\boldsymbol{c}_{m,n}^\tau$ with the set size being $N_m^\tau$.
	}
	\vspace{-1em}
	\label{fig:metaGraphicalRepresentation}
\end{figure}
Besides the limited annotations, early misinformation detection faces a new challenge that has not been considered in previous studies. At the early stage of misinformation emerging on social media, there is limited evidential content. Therefore, it is important to study the shared structure of how to utilize limited evidence.

Inspired by few-shot learning, we formulate the early detection of online misinformation as a new few-shot-few-clue learning framework. As the input of each datum is made up of a statement and its evidential clues, the concept of {few-shot-few-clue} describes scenarios of limited size of data in two levels: (1) the number of annotated data (consisting of labels, statements and their accompanied clues) and (2) the number of clues per statement. 
Like few-shot misinformation detection,
the concept ``few'' 
of few-clue indicates a specific number like 5-clue misinformation detection.
Few-shot-few-clue learning differs from conventional few-shot learning because it considers fewness at two levels: shot and clue inside each shot.

Figure~\ref{fig:metaGraphicalRepresentation} is a graphical representation of variables of interest. 
Similar to few-shot misinformation detection, the statement veracity $y$ is decided by a detection model with: (1) the parameter $\boldsymbol{\phi}$ and (2) the input of the statement content $\boldsymbol{s}$ and the corresponding clues $\boldsymbol{c}$.
What makes few-shot-few-clue learning different is the decomposition of the clue variable. The variable $\boldsymbol{c}_{m}^\tau$ is decomposed into a set of clues $\boldsymbol{c}_{m,n}^\tau$ with the set size being $N_m^\tau$. This is in line with our interest in cases when there are limited available clues per statement during the early stages, i.e.,the effect of $N_m^\tau$.
Formally, we give definitions of relevant terms with reference to Figure~\ref{fig:metaGraphicalRepresentation}.
%
\begin{definition}{}
	A \emph{task} $\tau$ is defined as the topic-centered detection of misinformation on the web.
\end{definition}
\begin{definition}{}
	A \emph{shot} $m\in\{1,\ldots,M^\tau\}$ is defined as a data instance consisting of a label $y$, a statement $\boldsymbol{s}$ and its set of evidential clues $\{
	\boldsymbol{c}\}$. $M^\tau$ denotes the number of shots in the task $\tau$.
\end{definition}
\begin{definition}{}
	A \emph{clue} $\boldsymbol{c}$ is defined as evidence that contributes to misinformation detection. $N_m^\tau$ denotes the number of clues in the shot $m$ of the task $\tau$.
\end{definition}
We make three further declarations: (1) misinformation about different events/topics {is} mapped to different tasks; 
(2) a statement is in the form of text; (3) a clue is an user engagement.
Based on the above definitions, the research problem of few-shot-few-clue applies when there are few statements (small $M^\tau$) and few clues per statement (small $N_m^\tau$) for the early detection of misinformation regarding.

The meta-parameter $\boldsymbol{\theta}$ can be interpreted as the common patterns shared by detection models across different topics; 
while the parameter $\boldsymbol{\phi}$ can be interpreted as features specific to certain topics.
For early detection when the size of $\mathcal{D}^{\mathcal{T}}$ is limited, machine learning models usually perform poorly on $\tilde{\mathcal{D}}^{\mathcal{T}+1}$; the meta-knowledge contained in $\boldsymbol{\theta}$ can be helpful.

\begin{figure*}[t]
	\center
	\includegraphics[width=0.8\textwidth]{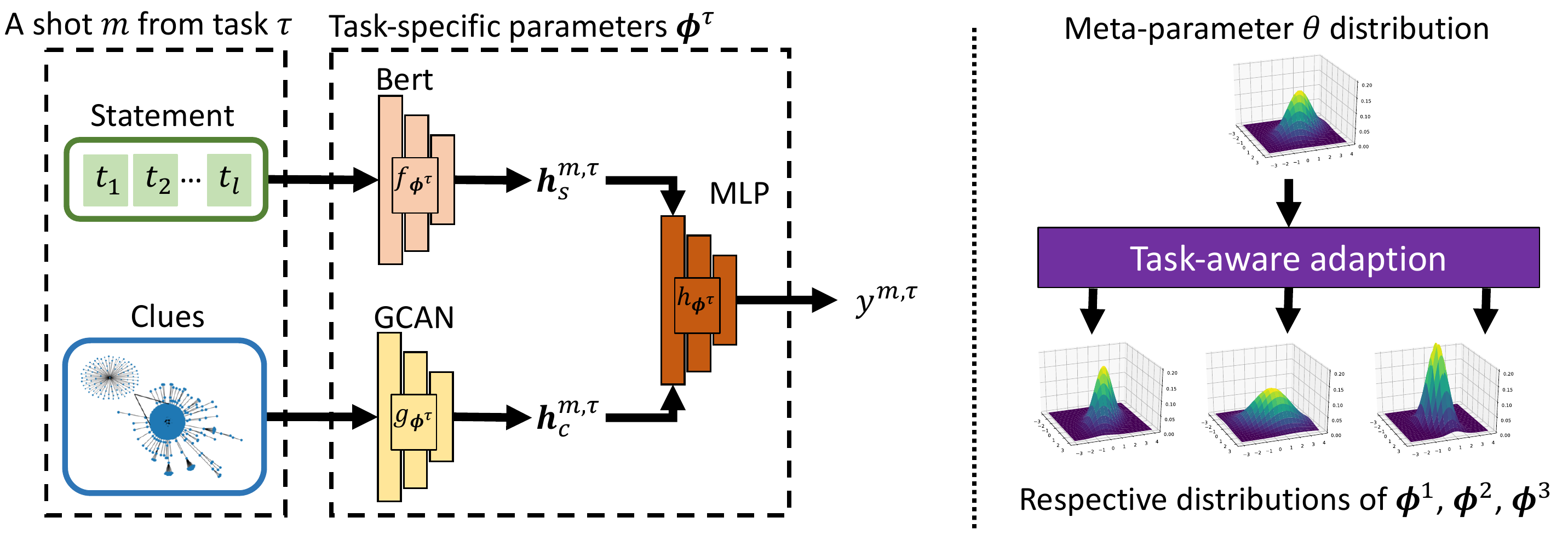}
	\caption{The left is a schematic illustration of the base model and the right shows the meta-parameter distribution is converted through task-aware adaption into respective task-specific parameter distributions.
	}
	\vspace{-1em}
	\label{fig:base_model}
\end{figure*}

\section{Methodology}
\label{sec:method}
After formulating early misinformation detection as a few-shot-few-clue learning framework, we propose a {task-aware} Bayesian meta-learning algorithm as a solution. 

\subsection{Base Model}
\noindent \paragraph{\textbf{Content feature extractor}}
We employ the pre-trained Bert model proposed by~\citet{devlin-etal-2019-bert} to extract text features. The input to Bert is a sequence of word tokens $(t_1, \cdots, t_l)$ and the output is a text embedding that summarizes linguistic and semantic features. We chose the base version of the pre-trained Bert for the sake of computation complexity; a fully connected layer converts the Bert output to a lower-dimension embedding. 
We freeze the parameters of Bert to avoid overfitting; only the fully connected layer is counted in the trainable parameters $\boldsymbol{\phi}$.
%
For a task $\tau$ with $M^\tau$ data points, the statement of the $m$-th ($1\leq m\leq M^\tau$) datum is input to
a function $f_{\boldsymbol{\phi}^\tau}(\cdot)$ to obtain the embedding vector:
\begin{equation}
\boldsymbol{h}_s^{m,\tau} = f_{\boldsymbol{\phi}^\tau}(\boldsymbol{s}_m^\tau).
\end{equation}

\noindent \paragraph{\textbf{Social context encoder}}
We consider a set of clues in the temporal order $\boldsymbol{c}_m^\tau=\left\{\boldsymbol{c}_{m,1}^\tau,\ldots,\boldsymbol{c}_{m,n}^\tau,\ldots, \boldsymbol{c}_{m,N}^\tau\right\}$, 
where $\boldsymbol{c}_{m,n}^\tau$ denotes a clue. 
For the social context encoder, we use the state-of-the-art GCAN~\cite{lu2020gcan} model that extracts user features from their profiles and social engagements, and uses CNN and RNN to generate embeddings of repost propagation. A graph is constructed to describe the engagements between users, and a GCN is used to learn the graph representation. 
We use the function $g_{\boldsymbol{\phi}^\tau}(\cdot)$ to denote this encoder:
\begin{equation}
\boldsymbol{h}_{\boldsymbol{c}}^{m,\tau} = g_{\boldsymbol{\phi}^\tau}(\boldsymbol{c}_m^\tau).
\end{equation}
The output is a concatenation of source-interaction co-attention embeddings, source-propagation embeddings, and propagation graph embeddings. 
All trainable GCAN parameters are counted in $\boldsymbol{\phi}$.

\noindent \paragraph{\textbf{Aggregator}}
After obtaining $\boldsymbol{h}_s^{m,\tau}$ and $\boldsymbol{h}_{\boldsymbol{c}}^{m,\tau}$, we investigate how to incorporate clues to predict statement veracity.
Veracity is predicted by combining the statement and clue information via an aggregator $h_{\boldsymbol{\phi}^\tau}(\cdot,\cdot)$. 
We implement it as a multi-layer perceptron (MLP).
Basically, $h_{\boldsymbol{\phi}^\tau}(\cdot,\cdot)$ takes the concatenation of the statement embedding and the social context embedding as inputs, 
\begin{equation}
y^{m,\tau} =h_{\boldsymbol{\phi}^\tau}(\boldsymbol{h}_s^{m,\tau},\boldsymbol{h}_{\boldsymbol{c}}^{m,\tau}).
\end{equation}
All of the three modules $f_{\boldsymbol{\phi}^\tau}(\cdot)$, $g_{\boldsymbol{\phi}^\tau}(\cdot)$ and $h_{\boldsymbol{\phi}^\tau}(\cdot,\cdot)$ are parameterized by the topic-specific parameter $\boldsymbol{\phi}^\tau$. Trainable MLP parameters are counted in $\boldsymbol{\phi}$.

\subsection{{Task-Aware} Bayesian Meta-Learning}

To efficiently learn the model parameters, we design a {task-aware} Bayesian meta-learning algorithm. 
%
One concern of meta-learning's application to misinformation detection is that tasks might be dissimilar and meta-parameters are not transferable. To address this concern, we introduce {topic similarity} $Sim(\cdot,\cdot)$ to compute semantic similarity between topics, which determines how much we rely on the meta-parameter $\boldsymbol{\theta}$ to learn the task-specific parameter $\boldsymbol{\phi}^\tau$. 
We expect $\boldsymbol{\phi}^\tau$ depends heavily on $\boldsymbol{\theta}$ when a topic $\tau$ is similar to topics in the meta-training set.

Specifically, we compute a topic embedding $\boldsymbol{E}^{\tau}$ using a word  embedding $\boldsymbol{e}_i$ (based on Word2Vec~\cite{mikolov2013efficient}) and its weight $w_i$ (the normalized frequency of a word in the topic),
where $i$ indicates the $i$-th word in the vocabulary $V$:
\begin{equation}
\boldsymbol{E}^{\tau} = \sum_{i \in \{1,\cdots,|V|\}}{w_i * \boldsymbol{e}_i}.
\end{equation}
For all topics in the meta-training set, we compute an embedding:
\begin{equation}
\boldsymbol{E}_{\rm{meta-train}}=\sum_{\tau \in \{1, \cdots,\mathcal{T}\} }{\frac{|\mathcal{D}^\tau|}{|\mathcal{D}_{\rm{meta-train}}|}\boldsymbol{E}^{\tau}}
\end{equation}
where $|\mathcal{D}^\tau|$ and $|\mathcal{D}_{\rm{meta-train}}|$ are the sizes of the dataset $\mathcal{D}^\tau$ and the meta-training set respectively.
To avoid overfitting of $\boldsymbol{\phi}^\tau$ for a topic $\tau$, we introduce an adaptive parameter $z^{\tau}$ based on topic similarity:
\begin{equation}
z^{\tau} = \rm{cos}(\boldsymbol{E}^{\tau}, \boldsymbol{E}_{\rm{meta-train}}),
\end{equation}
where $\rm{cos}(\cdot,\cdot)$ is the cosine function.
Then we use $z^{\tau}$ to determine the initialization of the task-specified model parameter $\boldsymbol{\phi}^{\tau}$:
\begin{equation}
\boldsymbol{\phi}^{\tau}_{\rm{init}}=z^{\tau}\boldsymbol{\theta}+(1-z^{\tau})\boldsymbol{\theta}_{\rm{rand}}
\label{eq:semantic_theta}
\end{equation}
where $\boldsymbol{\theta}_{\rm{rand}}$ is randomly sampled from a normal distribution. When $\tau$ is more similar to meta-training set, $\boldsymbol{\phi}^{\tau}_{\rm{init}}$ depends more on $\boldsymbol{\theta}$ learnt from the meta-training set, otherwise there will be more randomness brought by $\boldsymbol{\theta}_{\rm{rand}}$.

{
The task-aware Bayesian meta-learning algorithm can provide two benefits: (1) uncertainty estimation that is infeasible for alternative optimization-based meta-learning including MAML, and is less prone to be under-fitting than neural processes; and (2) topic similarity when adapting global knowledge to tasks, which addresses the concern of dissimilar task distributions in~\cite{wangmulti}. 
}



\clearpage
\section{Optimization}
\label{sec:optimization}

\subsection{Evidence Lower Bound}
To optimize the {task-aware} Bayesian meta-learning algorithm with variational inference, we first need to derive the Evidence Lower BOund (ELBO) of the data likelihood:
\begin{align}
& \mathcal{L}\left(\psi, \{\boldsymbol{\lambda}^{\tau}| \tau= 1, \ldots, \mathcal{T}\}\right) \nonumber \\
=&  \mathbb{E}_{q(\boldsymbol{\theta};\psi)} \left[\sum_{\tau=1}^\mathcal{T} \mathbb{E}_{q(\boldsymbol{\phi}^{\tau};\boldsymbol{\lambda}^{\tau})} \left[ \log p(\mathbf{Y}^{\tau}| \mathbf{S}^{\tau},\mathbf{C}^{\tau}, \boldsymbol{\phi}^{\tau} )   \right] \right. \nonumber\\ & \left. - D_{KL}(q(\boldsymbol{\phi}^{\tau};\boldsymbol{\lambda}^{\tau}) || p(\boldsymbol{\phi}^{\tau} | \boldsymbol{\theta})) \vphantom{\sum_{m=1}^{M^\tau}}
\right] 
- D_{KL}(q(\boldsymbol{\theta};\psi) || p(\boldsymbol{\theta})). 
\label{eqn:ELBO}
\end{align}
Derivation of Eq.~\ref{eqn:ELBO} can be found in Appendix~\ref{appendix:ELBO}.
Here, $\psi$ and $\{\boldsymbol{\lambda}^{\tau}| \tau= 1, \ldots, \mathcal{T}\}$ are the variational parameters (such as a mean and standard deviation of each weight) of the approximate posteriors over the global latent variables $\boldsymbol{\theta}$ and the local latent variables $\boldsymbol{\phi}^{\tau}$.
%
If we maintain distinct variational parameters $\boldsymbol{\lambda}^{\tau}$, each of which indexes a distribution over task-specific model parameters $q(\boldsymbol{\phi}^{\tau};\boldsymbol{\lambda}^{\tau})$, the number of variational parameters is in the scale of $O(1+\mathcal{T})$.
\begin{figure}[!t]
	\center
	\includegraphics[width=0.8\columnwidth]{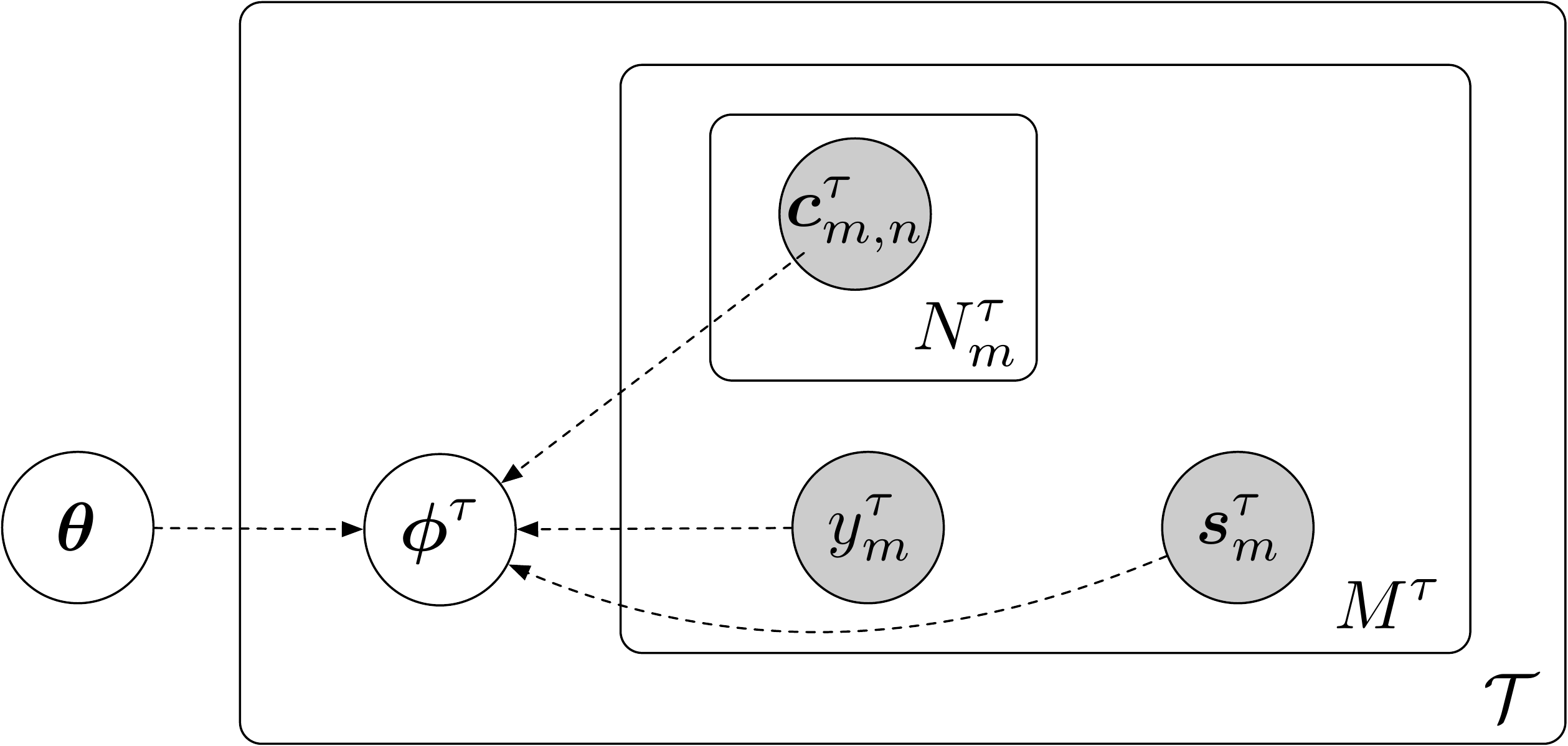}
	\caption{Dashed lines denote the inference process that obtains variational approximation to the intractable posterior.
	}
	\vspace{-1em}
	\label{fig:metaGraphicalRepresentation_infer}
\end{figure}

\subsection{Scaling Meta-Learning with Amortized Variational Inference}
\begin{algorithm}[!t]
	\SetKwInOut{Input}{input}\SetKwInOut{Output}{output}
	\caption{The task-aware Bayesian meta-training procedure.}
	\label{alg:metalearning}
	\Input{The lower and upper bound of update steps $K_{\text {min}} $ and $K_{\text {max}} $, the number of total tasks $\mathcal{T}$, the inner and outer learning rate $\alpha$ and $\beta$, the topic embedding in the meta-training set $\boldsymbol{E}_{\rm{meta-train}}$
	}
	\Begin{
		Initialize $\boldsymbol{\theta} = \{\boldsymbol{\mu}_{\boldsymbol{\theta}},\boldsymbol{\sigma}_{\boldsymbol{\theta}}^2\}$
		
		$p(\theta)=\mathcal{N}(\boldsymbol{\mu} ; \mathbf{0}, \mathbf{I}) \cdot \prod_{l=1}^{D} \operatorname{Gamma}\left(\gamma_l ; a_{0}, b_{0}\right)$
		
		\For{$\tau$ from 1 to $\mathcal{T}$}{
			
			$\mathcal{D}^{\tau}=\left\{{S}^{\tau},{C}^{\tau}, {Y}^{\tau}\right\}$
			
			$\boldsymbol{E}^{\tau} = \sum_{i \in \{1,\cdots,|V|\}}{w_i * \boldsymbol{e}_i}$
			
			$z^{\tau} = \rm{cos}(\boldsymbol{E}^{\tau}, \boldsymbol{E}_{\rm{meta-train}})$
			
			
			$\boldsymbol{\mu}_{\boldsymbol{\lambda}}^{(0)} \leftarrow z^{\tau}\boldsymbol{\mu}_{\boldsymbol{\theta}}+(1-z^{\tau})\boldsymbol{\mu}_{\boldsymbol{\theta}}^{\rm{rand}}$
			
			$\boldsymbol{\sigma}^{2 {(0)}}_{\boldsymbol{\lambda}} \leftarrow z^{\tau}\boldsymbol{\sigma}_{\boldsymbol{\theta}}^{2}+(1-z^{\tau})\boldsymbol{\sigma}_{\boldsymbol{\theta}}^{2~\rm{rand}}$
			
			
			$K^{\tau}= \ceil*{z^{\tau} K_{\text {min}} + (1-z^{\tau}) K_{\text {max}}}$
			
			\For{k=0 to $K^{\tau}-1$}{
				$\boldsymbol{\lambda}^{(k)} \leftarrow\left\{\boldsymbol{\mu}_{\boldsymbol{\lambda}}^{(k)}, \boldsymbol{\sigma}_{\boldsymbol{\lambda}}^{2(k)}\right\}$
				
				$\boldsymbol{\mu}_{\boldsymbol{\lambda}}^{({k}+1)} \leftarrow \boldsymbol{\mu}_{\boldsymbol{\lambda}}^{({k})}-\alpha \nabla_{\boldsymbol{\mu}_{\boldsymbol{\lambda}}^{(k)}} \mathcal{L}_{\mathcal{D}^{\tau}}\left(\boldsymbol{\lambda}^{(k)}, \boldsymbol{\theta}\right)$
				
				$\boldsymbol{\sigma}^{2 {({k}+1)}}_{\boldsymbol{\lambda}} \leftarrow \boldsymbol{\sigma}^{2({k})}_{\boldsymbol{\lambda}} -\alpha \nabla_{\boldsymbol{\sigma}^{2({k})}_{\boldsymbol{\lambda}}} \mathcal{L}_{\mathcal{D}^{\tau}}\left(\boldsymbol{\lambda}^{(k)}, \boldsymbol{\theta}\right)$
			}
			
			$\boldsymbol{\lambda}^{(K^{\tau})} \leftarrow\left\{\boldsymbol{\mu}_{\boldsymbol{\lambda}}^{(K^{\tau})}, \boldsymbol{\sigma}_{\boldsymbol{\lambda}}^{2(K^{\tau})}\right\}$
			
			$q(\boldsymbol{\theta})=\mathbbm{1}\left\{\boldsymbol{\mu}=\boldsymbol{\mu}_{\theta}\right\} \cdot \mathbbm{1}\left\{\boldsymbol{\sigma}^{2}=\boldsymbol{\sigma}_{\boldsymbol{\theta}}^{2}\right\}$
			
			$\boldsymbol{\mu}_{\boldsymbol{\theta}} \leftarrow \boldsymbol{\mu}_{\boldsymbol{\theta}}-\beta \nabla_{\boldsymbol{\mu}_{\boldsymbol{\theta}}}\left[\mathcal{L}_{\mathcal{D}^\tau}\left(\boldsymbol{\lambda}^{(K)}, \boldsymbol{\theta}\right)+\frac{1}{\mathcal{T}} \operatorname{D_{KL}}(q(\boldsymbol{\theta}) \| p(\boldsymbol{\theta}))\right]$
			
			$\boldsymbol{\sigma}_{\boldsymbol{\theta}}^2 \leftarrow \boldsymbol{\sigma}_{\boldsymbol{\theta}}^2 -\beta \nabla_{\boldsymbol{\sigma}_{\boldsymbol{\theta}}^2 }\left[\mathcal{L}_{\mathcal{D}^\tau}\left(\boldsymbol{\lambda}^{(K)}, \boldsymbol{\theta}\right)+\frac{1}{\mathcal{T}} \operatorname{D_{KL}}(q(\boldsymbol{\theta}) \| p(\boldsymbol{\theta}))\right]$ 
		}
		
		\Return ${\theta}$
	}
\end{algorithm}

To scale the optimization process, 
we adopt the Amortized Variational Inference (AVI) for meta-learning proposed by~\citet{ravi2018amortized} to infer the posterior distributions of $\boldsymbol{\phi}^{\tau}$ and $\boldsymbol{\theta}$.
As shown by Model-Agnostic Meta-Learning~\citep{finn2017model}, a proper initialization can produce the decent task-specific parameters. Hence, from a global initialization $\boldsymbol{\theta}$, we produce the variational parameter $\boldsymbol{\lambda}^{\tau}$ by conducting several steps of gradient descent. 
%

First, we need to specify the loss function on $\mathcal{D}^{\tau}=\left\{{S}^{\tau},{C}^{\tau}, {Y}^{\tau}\right\}$,
\begin{align}
\mathcal{L}_{D^\tau}(\boldsymbol{\lambda}, \boldsymbol{\theta})  = & \mathbb{E}_{q(\boldsymbol{\phi}^{\tau};\boldsymbol{\lambda})} \left[ \log p(\mathbf{Y}^{\tau}| \mathbf{S}^{\tau},\mathbf{C}^{\tau}, \boldsymbol{\phi}^{\tau} )  \right] \nonumber\\
&- D_{KL}(q(\boldsymbol{\phi}^{\tau};\boldsymbol{\lambda}) || p(\boldsymbol{\phi}^{\tau} | \boldsymbol{\theta})).
\end{align}

Then, we modify the procedure of stochastic gradient descent, $SGD_{K^{\tau}}(\mathcal{D}^{\tau}, \boldsymbol{\theta})$, based on the adaptive parameter $z^{\tau}$, to produce $\boldsymbol{\lambda}^{\tau}$ from the the global initialization $\boldsymbol{\theta}$:
\begin{equation}
\begin{array}
{l}{\text { 1. } \boldsymbol{\lambda}^{(0)}=z^{\tau}\boldsymbol{\theta}+(1-z^{\tau})\boldsymbol{\theta}_{\rm{rand}}}\\
{\text { 2. } K^{\tau}= \ceil*{ z^{\tau} K_{\text {min}} + (1-z^{\tau}) K_{\text {max}} } }\\ 
{\text { 3. for } k=0, \ldots, K^{\tau}-1, \text { set }} \\ 
{\quad \boldsymbol{\lambda}^{(k+1)}=\boldsymbol{\lambda}^{(k)}-\alpha \nabla_{\boldsymbol{\lambda}^{(k)}} \mathcal{L}_{\mathcal{D}^{\tau}}\left(\boldsymbol{\lambda}^{(k)},  \boldsymbol{\theta}\right)} \\
{\text { 4. } \boldsymbol{\lambda}^{\tau} = \boldsymbol{\lambda}^{(K^{\tau})}}
\end{array}
\end{equation}
$\alpha$ is the learning rate and {$K^{\tau}$ is the number of steps of gradient descents that is determined by the adaptive parameter $z^{\tau}$. $K_{\text {min}}$ and $K_{\text {max}} $ denote the lower and upper bounds of $K^{\tau}$, and $\ceil \cdot$ is a ceil function.}
 Hereby
$q(\boldsymbol{\phi}^{\tau};\boldsymbol{\lambda}^{\tau})=q(\boldsymbol{\phi}^{\tau};SGD_{K^{\tau}}(\mathcal{D}_t^{\tau}, \boldsymbol{\theta}))=q(\boldsymbol{\phi}^{\tau}|\mathcal{D}^{\tau}, \boldsymbol{\theta})$.
With this form of the variational distribution, $\boldsymbol{\theta}$ serves as both the global initialization of local variational parameters $\boldsymbol{\lambda}$ and the parameters of the prior $p(\boldsymbol{\phi} | \boldsymbol{\theta})$. 

Finally, we estimate $\boldsymbol{\theta}$.
We let $q(\boldsymbol{\theta}; \psi)$ be a dirac delta function $q(\boldsymbol{\theta}) = \mathbbm{l}\{\boldsymbol{\theta} = \boldsymbol{\theta}^{\star}\}$, where $\boldsymbol{\theta}^{\star}$ is the solution to the optimization problem. This conduction removes the need for global variational parameters $\psi$ and simplifies the optimization problem of Eq.~\ref{eqn:ELBO} to
\begin{align}
\label{eqn:obj_func}
\underset{\boldsymbol{\theta}}{\arg \max} \quad & \left[\sum_{\tau=1}^\mathcal{T} \mathbb{E}_{q(\boldsymbol{\phi}^{\tau}|D^\tau,\boldsymbol{\theta})} \left[\log p(\mathbf{Y}^{\tau}| \mathbf{S}^{\tau},\mathbf{C}^{\tau}, \boldsymbol{\phi}^{\tau} ) \right] \right. \nonumber\\
& \left. 
- D_{KL}(q(\boldsymbol{\phi}^{\tau}|D^\tau,\boldsymbol{\theta}) || p(\boldsymbol{\phi}^{\tau} | \boldsymbol{\theta})) \vphantom{\sum_{m=1}^{M^\tau}} \right] \nonumber\\
& - D_{KL}(q(\boldsymbol{\theta}) || p(\boldsymbol{\theta})).
\end{align}
%
%
Algorithm~\ref{alg:metalearning} provides the pseudo code for the {task-aware} Bayesian meta-training algorithm.

\begin{table}[h]
	\caption{Statistics of the used datasets.}
	\vspace{-1em}
	\label{tab:datasets}
	\resizebox*{0.9\columnwidth}{!}{
		\begin{tabular}{lrrr}
			\toprule
			Dataset               & Twitter 15 & Twitter 16 & Pheme \\ \midrule
			\# Source Tweets        & 742         & 412   &   5,152  \\
			\# True                 & 372         & 205    &   3,142 \\
			\# False                & 370         & 207    &   2,010 \\
			\# Users                & 190,868     & 115,036 &  36,647 \\
			avg. retweets per story & 292.19      & 208.70  &  64.24 \\ \bottomrule
	\end{tabular}}
	\vspace{-1em}
\end{table}

\begin{figure*}
	\centering
	\includegraphics[width=0.6\textwidth]{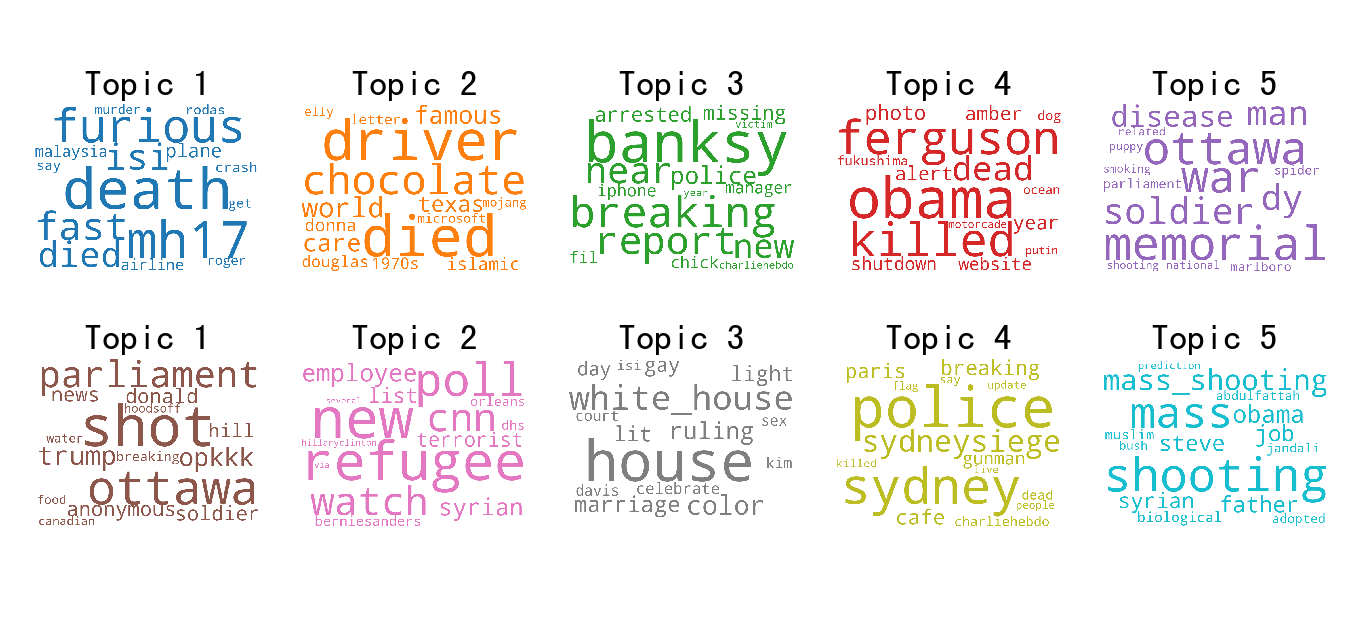}
	\vspace{-3.5em}
	\caption{
		Word cloud of topic clusters on  Twitter15 (the first row) and Twitter16 (the second row).}
	\label{fig:topic_cluster}
\end{figure*}

\section{Experiments}
\label{sec:experiments}
This section presents the experiments to evaluate the effectiveness of our {task-aware} Bayesian meta-learning (TABML~\footnote{The source code of TABML is publicly available from: \url{https://github.com/BML2021/bml}.}) algorithm. The research questions that guide the remainder of the paper are:
\begin{itemize}
	\item[RQ1] How is the misinformation detection performance of the proposed TABML algorithm compared to state-of-the-art misinformation detection baselines when full life-cycle data are available?
	\item[RQ2] What is the effect of the number of annotated statements on the performance of TABML and baselines in the setting of few-shot misinformation detection?
	\item[RQ3] What is the effect of the number of clues per statement on the performance of TABML and baselines in the setting of few-clue misinformation detection?
	\item[RQ4] Ablation study: what is the influence of the variable $z^{\tau}$ on the model's performance?
\end{itemize}
\begin{table*}[]
	\centering
	\caption{Performance comparison of the proposed Bayesian meta-learning algorithm against the baselines.}
	\vspace{-1em}
	\label{tab:comparison}
	\resizebox*{0.85\textwidth}{!}{
		\begin{tabular}{lccc|ccc|ccc}
			\toprule
			\multirow{2}{*}{Model} & \multicolumn{3}{c|}{Twitter15} & \multicolumn{3}{c}{Twitter16} & \multicolumn{3}{c}{Pheme}\\ 
			& Rec(\%) & Prec(\%) & Acc(\%) & Rec(\%) & Prec(\%) & Acc(\%) & Rec(\%) & Prec(\%) & Acc(\%)\\ 
			\midrule
			RFC  & 61.29 & 57.18 & 61.85 & 65.87 & 73.15 & 59.20  & 56.27 & 61.14 &60.53 \\
			CRNN  & 53.05 & 52.96 & 66.19  & 64.33 & 65.19 &65.76  & 63.14 & 59.98 & 61.01 \\
			CSI & 68.67 & 69.91 & 65.37 & 63.09 & 63.21 & 76.12 & 64.66 & 62.33 & 63.12\\
			dFEND  & 66.11 & 65.84 & 65.83  & 63.84 & 63.65 & 76.16 & 61.45 & 67.55 & 68.33\\
			GCAN  & \textbf{82.99} & 81.37 & 81.41  & 73.47 & \underline{81.61} & \underline{82.52}  & 58.42 & 63.26 & \underline{70.41}\\
			\hline
			EANN  & 64.56 & \underline{82.51} & 71.51 & 65.23 &  {81.04} & 79.05  & 65.45 & 66.36 & 69.79\\
			MetaFEND & 78.35 & 81.51 & \underline{82.44} & \underline{76.58} & 73.14 & {79.11} & \underline{70.54} & \underline{73.24} & {70.04}\\
			\hline
			Ours & \underline{82.62$\pm$1.71} & \textbf{83.86$\pm$1.26} & \textbf{85.65$\pm$1.93} & \textbf{80.77$\pm$2.02} & \textbf{82.32$\pm$3.16} & \textbf{86.14$\pm$1.41}  & \textbf{71.22$\pm$1.21} & \textbf{76.22$\pm$1.47} & \textbf{75.65$\pm$2.43}\\ 
			\bottomrule
	\end{tabular}}
\end{table*}

For evaluation purposes, we use three well-established benchmark datasets, i.e., Twitter15, Twitter16~\cite{ma2018detect} and Pheme~\cite{zhang2019reply}. 
Statement veracity is manually annotated by fact-checkers. These datasets also provide rich social context on top of the original statements: (1) users that are involved in the propagation of statements, and (2) social engagements in the temporal order. Table~\ref{tab:datasets} summarizes statistics of these datasets. 
For datasets without split of topics/events, including Twitter15 and Twitter16, we use Tweeter-LDA~\cite{diao2012finding}, an LDA variant widely used for short and noisy tweets, to determine topic clusters as well as important words with their weights. The Pheme dataset has been split based on events by its builders, so we compute word frequencies per topic as their weights, and keep the most frequent 10 words.
Among these clusters, we assign $80\%$ of them into a meta-training, 
and the rest into the meta-test subset to ensure there is no data overlap in subsets.
From each cluster {in Twitter15 and Twitter16}, we randomly sample 6 true shots and 6 false shots to make a data batch;
the average number of clue in a shot is 30. This can be denoted as 2-way-6-shot-30-clue learning. {Similarly, the learning process on Pheme can be denoted as 2-way-6-shot-25-clue learning.}
We set the inner and outer learning rate $\alpha$ and $\beta$ to be $2e-3$ and $1e-3$, respectively.
The lower and upper bound of update steps $K_{\text {min}} $ and $K_{\text {max}} $ are 2 and 8 respectively, and the dropout rate is 0.6 to avoid overfitting. 


\subsection{Baselines}

We compare the TABML algorithm and the following state-of-the-art models.
RFC~\cite{kwon2017rumor} is a random forest with features from the source tweets and engaged user profiles. 
CRNN~\cite{liu2018early} combines convolutional and recurrent neural networks to extract features from engaged users and retweet texts. 
CSI~\cite{ruchansky2017csi} incorporates relevant articles and analyses group behaviour of engaged users. 
dEFEND~\cite{shu2019defend} uses a co-attention mechanism to study source claims and user features. 
GCAN~\cite{lu2020gcan} utilises the co-attention graph networks to encapsulate the propagation structure of heterogeneous data.
EANN~\cite{wang2018eann} contains a event discriminator that removes event-specific features and discovers the common pattern among events. 
{MetaFEND~\cite{wangmulti} is a meta neural process that combines meta-learning and neural processes to detect few-shot misinformation.}

\begin{figure*}
	\centering
	\subfigure[]{ 
		\includegraphics[width=0.3\textwidth]{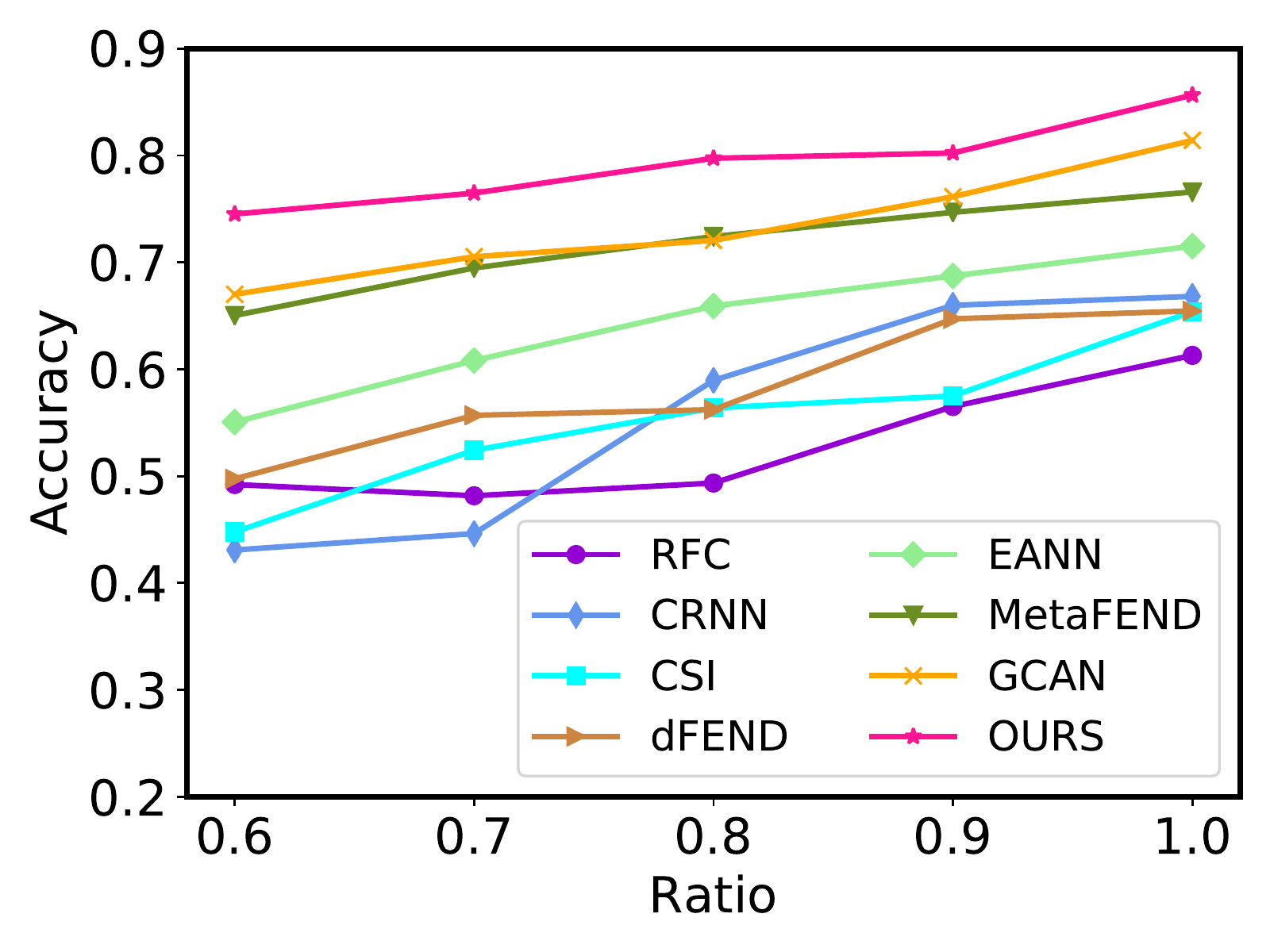}}
	\hspace{0in}
	\subfigure[]{
		\includegraphics[width=0.3\textwidth]{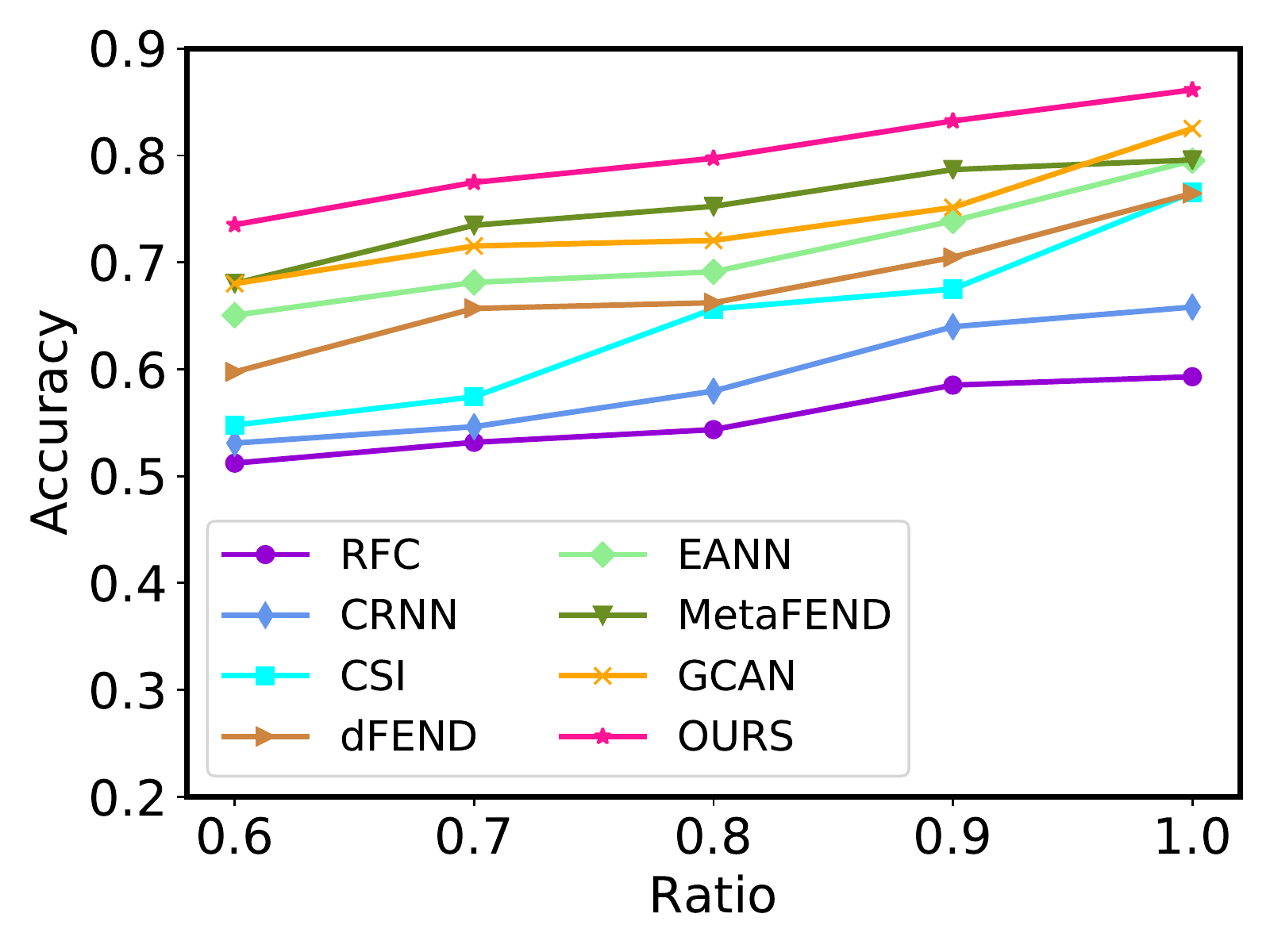}}
	\hspace{0in}
	\subfigure[]{
		\includegraphics[width=0.3\textwidth]{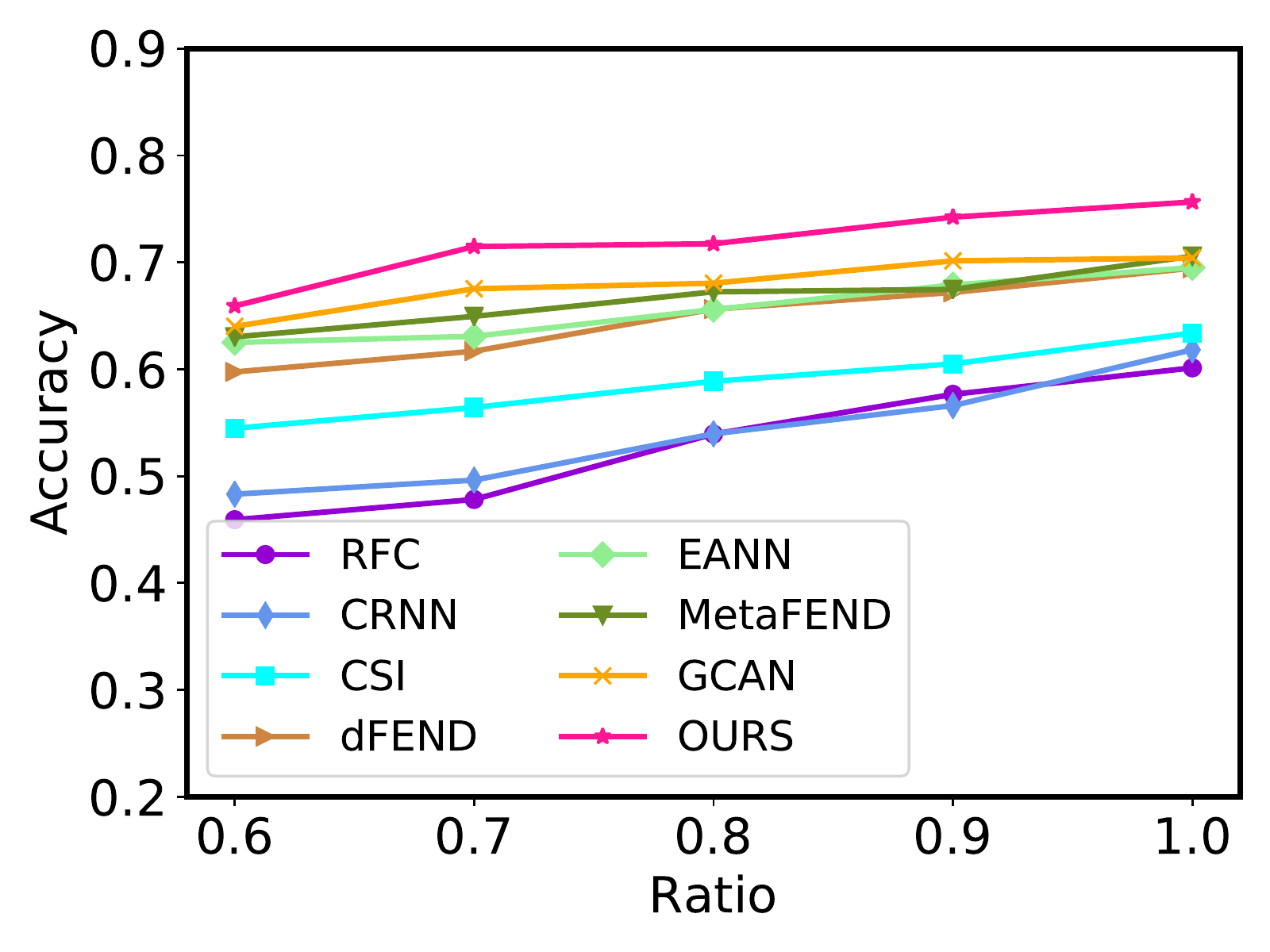}}
	\vspace{-1em}
	\caption{
		Accuracy of all models with different percentages of the training data on (a) Twitter15, (b) Twitter16 (b) and (c) Pheme.
	}
	\label{fig:comparison_few_shot}
\end{figure*}

\section{Results and Analysis}
\label{sec:results}

\subsection{Results of Topic Clustering}
We first examine the effectiveness of Twitter-LDA~\cite{diao2012finding} on determining the topic cluster that each tweet belongs to. 
{We present the results of topic clustering in the form of word cloud on Twitter15 and Twitter16 in Figure~\ref{fig:topic_cluster}.
} 
From this figure, it is clear that each topic cluster focuses well on a social event. For instance, the first topic of Twitter15 is about the ``MH17'' accident of Malaysia Airline, and the second topic of Twitter16 is more about politics. Moreover, there are sill overlap words between clusters, which indicates similarity in terms of semantics. This further verifies our assumption of similar tasks and serves as the basis of using semantic similarity to modify the Bayesian meta-learning based misinformation detection (that is, Equation~\ref{eq:semantic_theta}). 

The Pheme dataset has clearly split statement tweets into various events, hence there is no need to do topic clustering. These events are ``Charlie Hebdo'', ``Sydney siege'', ``Ferguson'', ``Ottawa shooting'', ``Germanwings-crash'', ``Putin missing'', ``Prince Toronto'', ``Gurlit'' and ``Ebola Essien''.

\subsection{Performances under all Available Statements and Clues (RQ1)}
\label{sec:RQ1}
Subsequently, we aim to evaluate the effectiveness of the proposed task-aware Bayesian meta-learning algorithm with all available statements and clues. Table~\ref{tab:comparison} summarizes the four evaluation metrics of all models on the three datasets. The reported numbers in Table~\ref{tab:comparison} are the means and standard deviations of TABML model running with different weights sampled from the same distribution that is learned at the meta-training phase. 
Our model outperforms all the baselines on most of all the metrics across the three datasets, achieving a performance improvement of around 4\% in terms of the accuracy. As for recall and precision, our model obtains the best values in most cases or the second best with a fairly minor gap in other cases.
On the larger Pheme dataset where state-of-the-art GCAN has severely discounted performance, our model enjoys a more significant improvement.
Compared to the baseline with transferable patterns, i.e., EANN and MetaFEND, our model outperforms them in terms of all metrics over three datasets. 
Therefore, our model is able to generalize better to new topics.

\begin{figure*}
	\centering
	\subfigure[]{ 
		\includegraphics[width=0.3\textwidth]{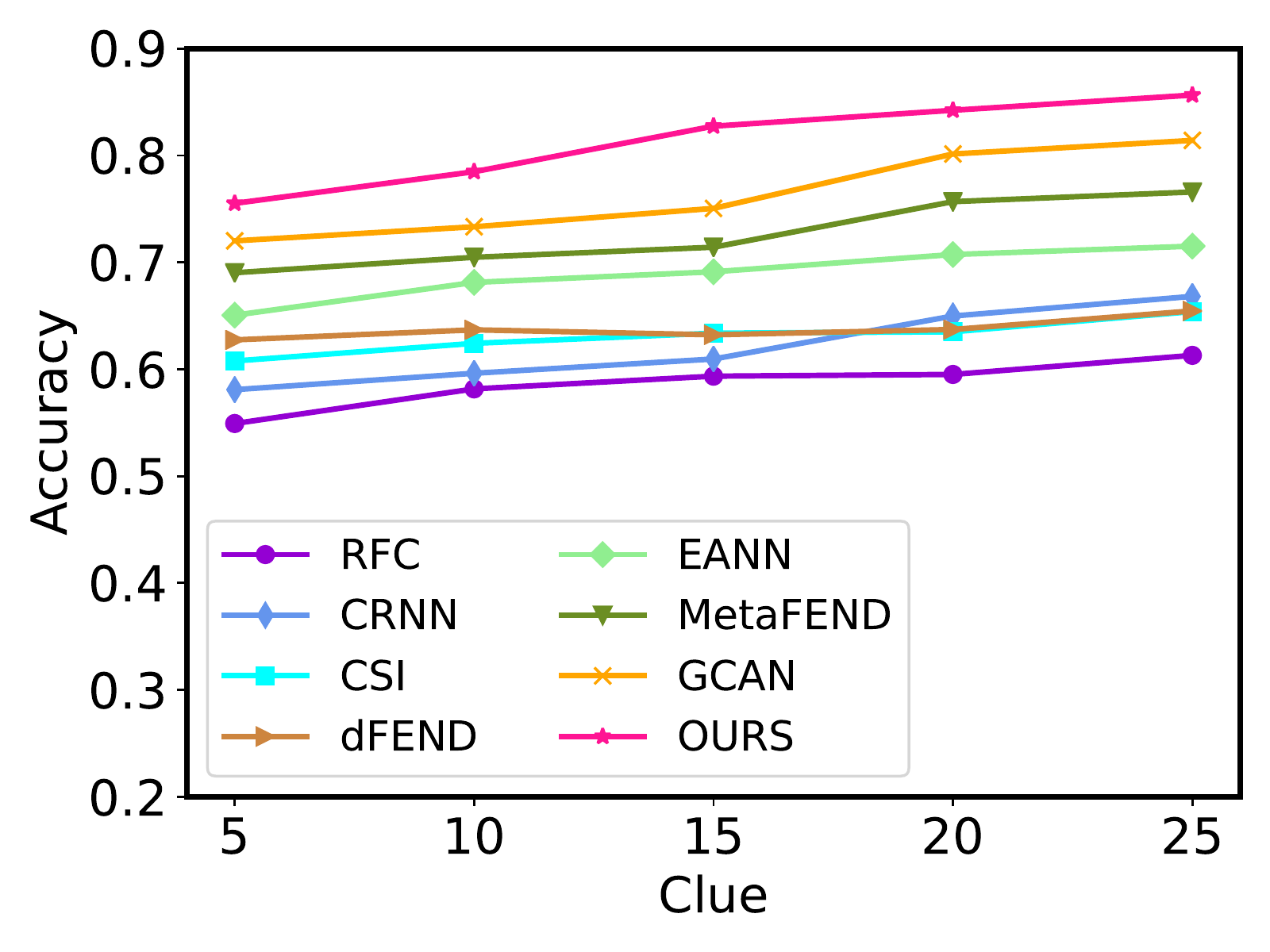}}
	\hspace{0in}
	\subfigure[]{
		\includegraphics[width=0.3\textwidth]{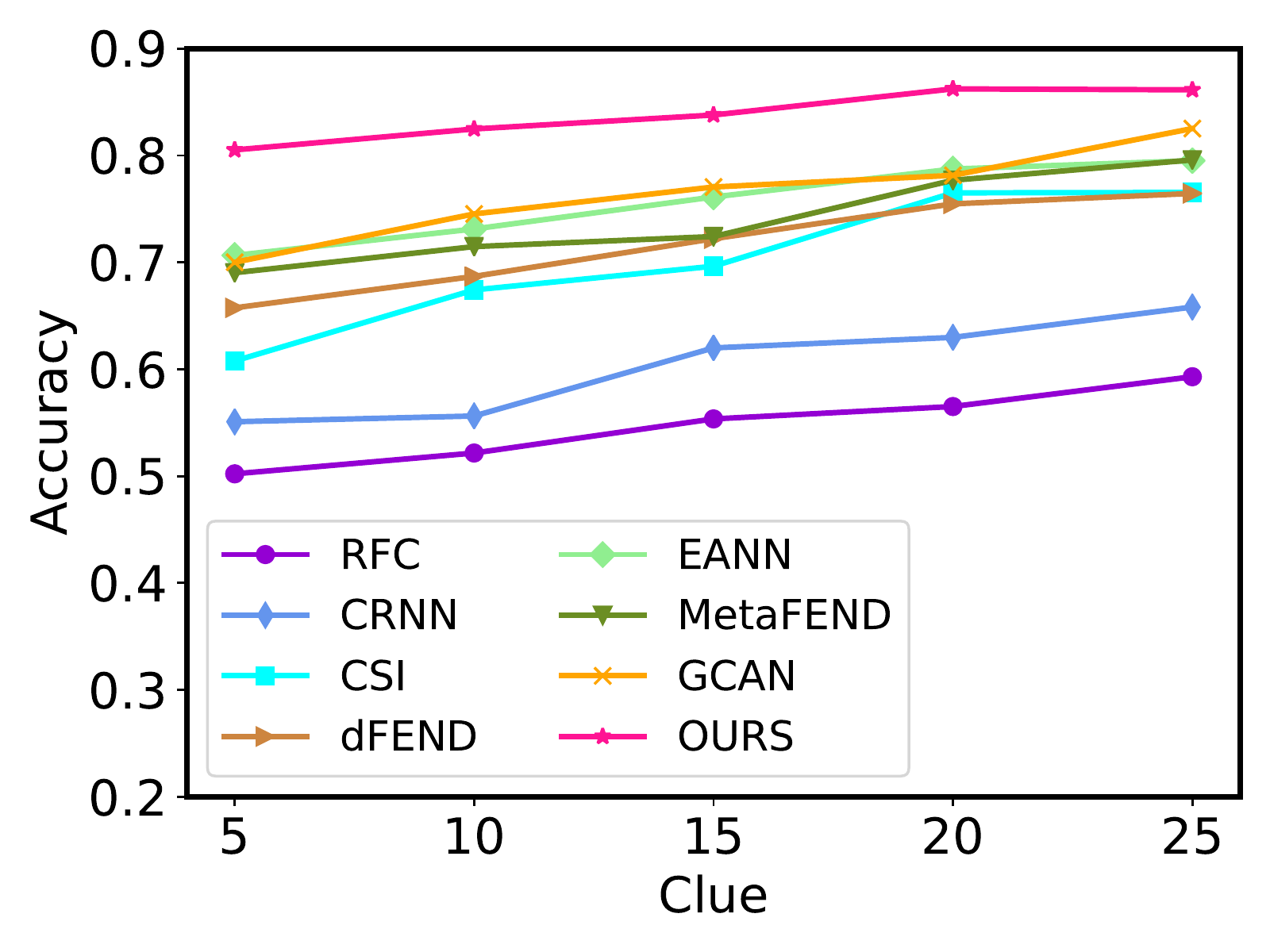}}
	\hspace{0in}
	\subfigure[]{
		\includegraphics[width=0.3\textwidth]{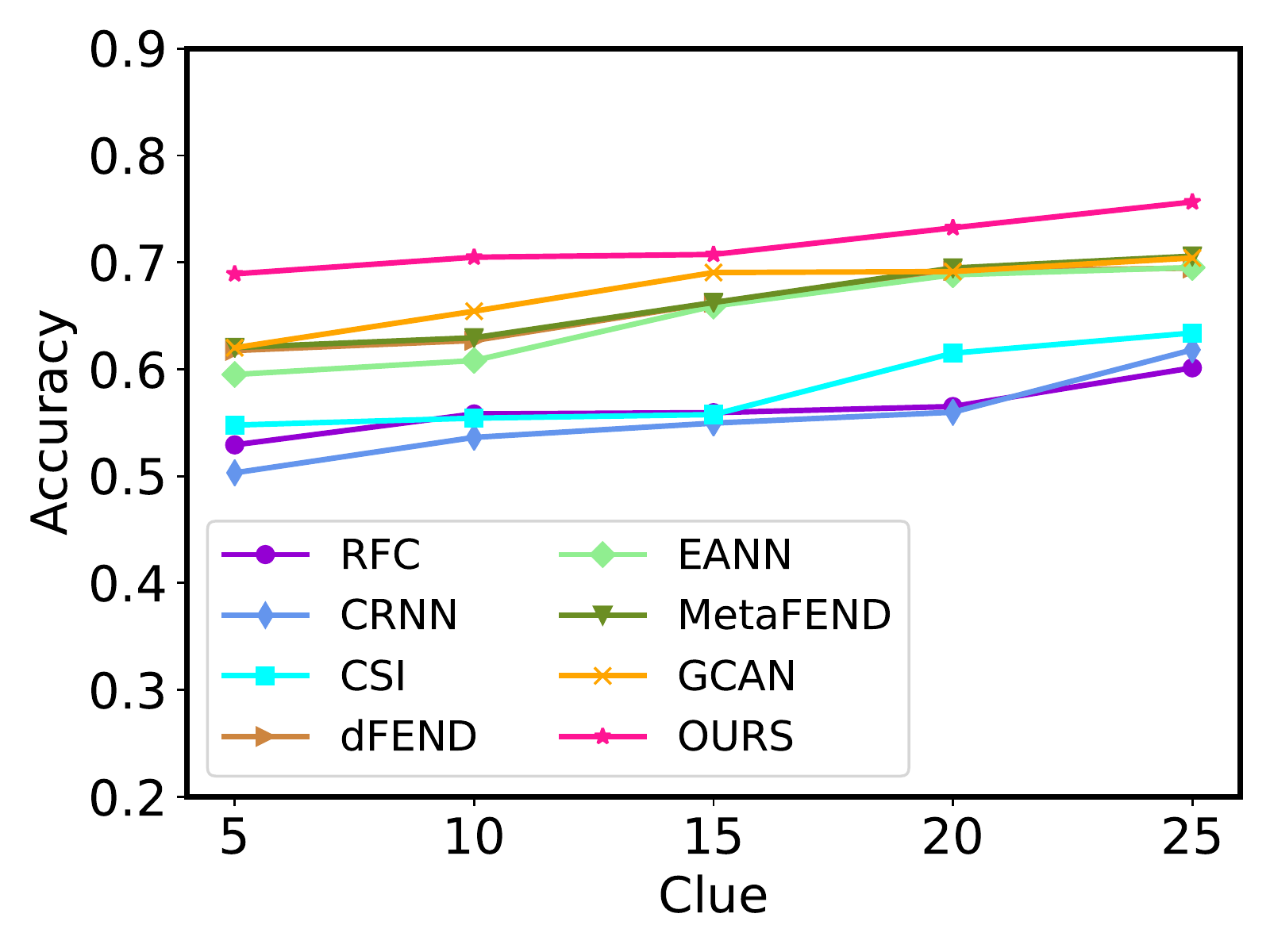}}
	\vspace{-1em}
	\caption{
		Accuracy of all models with different clues on (a) Twitter15, (b) Twitter16, and (c) Pheme.
	}
	\label{fig:comparison_few_clue}
\end{figure*}


\subsection{Performances under Limited Available Statements (RQ2)}
\label{sec:RQ2}
Annotating statement veracity is expensive and time-consuming. To understand the effect of the number of training statements, we conduct experiments using a different number of annotated statements. Specifically, we use five different percentages (i.e., 60\%, 70\%, 80\%, 90\% and 100\%) of all training data to train the models. From each topic cluster in the dataset {Twitter15 and Twitter16}, we still randomly sample 6 true and 6 false shots to make a batch, with each shot containing all available clues. So it is still 2-way-6-shot-30-clue learning {on Twitter15 and Twitter16. In the same way, it is 2-way-6-shot-25-clue learning on Pheme.}
Evaluation performance is reported in~Figure~\ref{fig:comparison_few_shot}.
We have three observations: First, it is clear that the size of training dataset exerts a significant effect on the performance of all models. Specifically, the performance increases as there are more training data in most cases.
Therefore, the most effective way to improve misinformation detection model is by increasing the number of annotated statements.
Second, our TABML is able to outperform the state-of-the-art baselines with different numbers of shots. Third, our TABML gains more stable performance improvement  compared with baselines when more training data are available.

\begin{figure}[!t]
	\centering
	\vspace{-1em}
	\includegraphics[width=0.8\columnwidth]{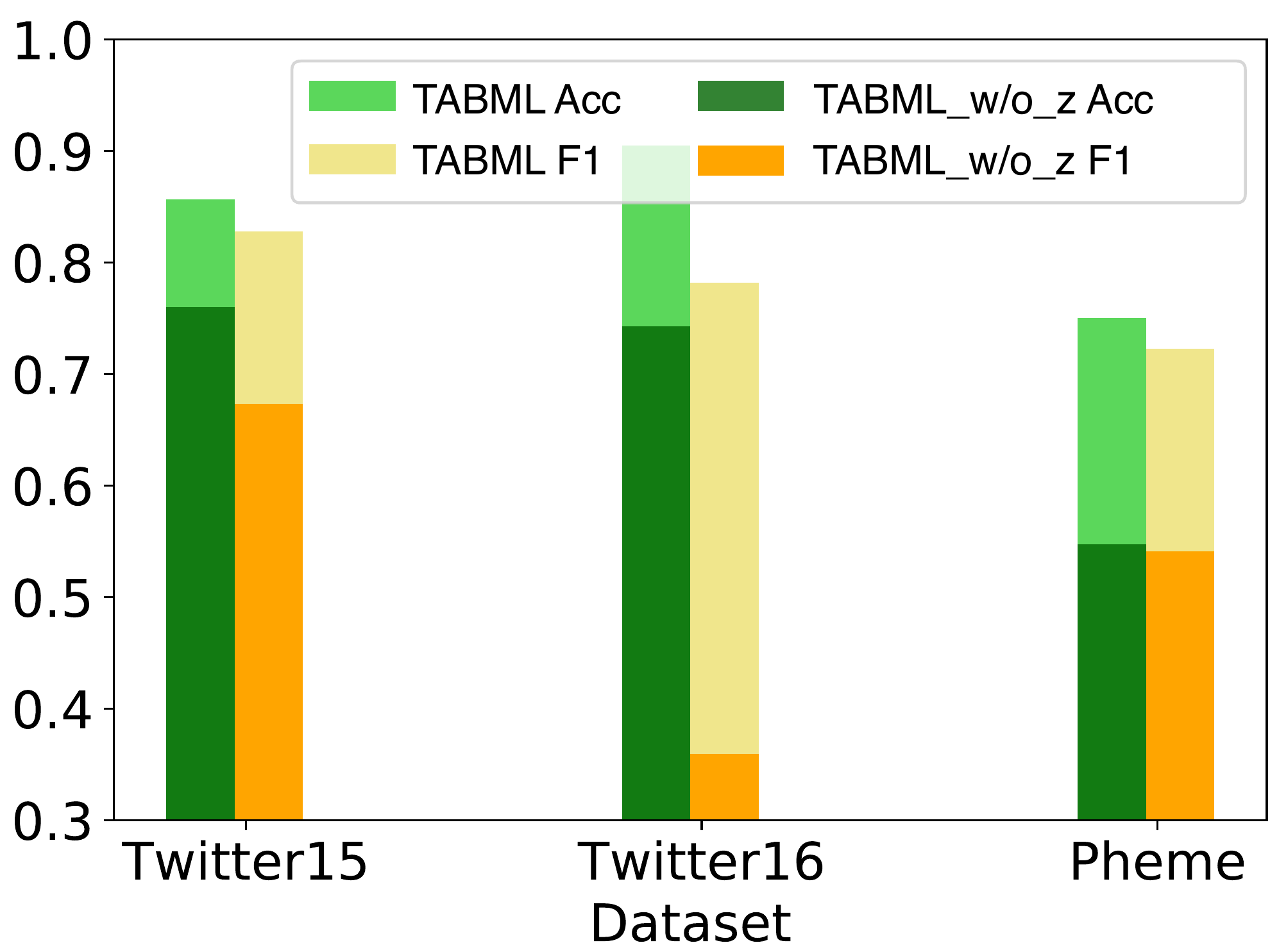}
	\vspace{-1em}
	\caption{Metrics of TABML with and without $\boldsymbol{z}$ on  Twitter15, Twitter16 and Pheme.}
	\label{fig:ablation}
	\vspace{-1em}
\end{figure}

\subsection{Performances under Few Clues (RQ3)}
\label{sec:RQ3}

Next, we study early misinformation detection when there are a limited number of clues per statement. From each topic cluster, {we fix the number of shots at 6. Then we change the number of clues $x$ in each shot, where $x$ can be [5, 10, 15, 20, 25, 30] and [5, 10, 15, 20, 25] for Twitter datasets and Pheme dataset respectively.} 
Figure~\ref{fig:comparison_few_clue} illustrates how the detection performance changes over different numbers of clues.
Obviously, our proposed TABML consistently shows better performance when varying the clues.
Another obvious finding is that, despite of the overall increasing accuracy with more clues, performance is improvement with different pace. This indicates that some clues hinder veracity prediction, and should be excluded. We leave this in the future work.
Baseline models show greater variances, which means instability and lower performance, while our TABML has more stability. In fact, Figure~\ref{fig:comparison_few_clue} shows TABML is stable enough to deal with few-clue problems.

\subsection{Ablation Study (RQ4)}
Finally, we study how the adaptive parameter $\boldsymbol{z}$ affects the performance of our model. Figure~\ref{fig:ablation} shows the comparative result. As we can see, our TABML model with the adaptive parameter $\boldsymbol{z}$ always outperforms the counterpart without $\boldsymbol{z}$ across three datasets no matter what metric it is, leading to $10\% \sim 30\%$ improvements. The result illustrates that TABML with  $\boldsymbol{z}$ can adaptively learn about the meta information among the topics and then effectively apply them into the learning of new tasks, avoiding the phenomenon of overfitting and the reduced performance in the learning of new tasks, which always contain the out of distributional training data.



\section{Conclusion}
\label{sec:conclusion}

In this paper, we have proposed TABML, a model for misinformation detection at early stages. We have investigated misinformation detection with limited statements and clues. Considering two levels of limited resources, we formulate a new few-shot-few-clue learning task. Solving this problem needs to recognize the shared patterns among different events and topics. We then develop a {task-aware} Bayesian meta-learning algorithm to discover such patterns.
%
An amortized variational inference method is derived for scalable parameter optimization. 
Experimental results confirm the superiority of our algorithm against the state-of-the-art methods for early misinformation detection. 
The findings highlight the importance of shared patterns among different events/topics and provide insights into misinformation detection in low-resource languages. 

	\appendix


\section{Appendix: Derivation of ELBO}
\label{appendix:ELBO}
The posterior distributions of meta-parameters and task-specific parameters in this paper are intractable.
To approximate the these posterior distributions, we resort to the variational inference (VI) method.
We present a detailed derivation of the objective function of Evidence Lower Bound (ELBO) in the VI method, i.e., Eq.~\ref{eqn:ELBO}.
\begin{equation}    
\begin{split}
\label{eqn:all_loss}
&\log \left[\prod_{\tau=1}^\mathcal{T} p(\mathbf{Y}^{\tau} | \mathbf{S}^{\tau},\mathbf{C}^{\tau})\right]  \\  %
=& \log \left[\prod_{\tau=1}^\mathcal{T} \int p(\mathbf{Y}^{\tau},  \boldsymbol{\phi}^{\tau}| \mathbf{S}^{\tau},\mathbf{C}^{\tau}) d\boldsymbol{\phi}^{\tau} \right] \nonumber \\
=& \log \left[\prod_{\tau=1}^\mathcal{T} \int p(\mathbf{Y}^{\tau}| \mathbf{S}^{\tau},\mathbf{C}^{\tau},  \boldsymbol{\phi}^{\tau}) p( \boldsymbol{\phi}^{\tau}) d\boldsymbol{\phi}^{\tau} \right] \nonumber \nonumber \\
=& \log \left[ \prod_{\tau=1}^\mathcal{T} \int p(\mathbf{Y}^{\tau}| \mathbf{S}^{\tau},\mathbf{C}^{\tau},  \boldsymbol{\phi}^{\tau}) \left(\int p( \boldsymbol{\phi}^{\tau},\boldsymbol{\theta})d\boldsymbol{\theta}\right) d\boldsymbol{\phi}^{\tau} \right] \nonumber \\
=& \log \left[\prod_{\tau=1}^\mathcal{T} \int p(\mathbf{Y}^{\tau}| \mathbf{S}^{\tau},\mathbf{C}^{\tau},  \boldsymbol{\phi}^{\tau}) \left(\int p( \boldsymbol{\phi}^{\tau}|\boldsymbol{\theta}) p(\boldsymbol{\theta}) d\boldsymbol{\theta}\right) d\boldsymbol{\phi}^{\tau} \right] \nonumber \\
= & \log \left[ \int p( \boldsymbol{\theta} ) \left( \prod_{\tau=1}^\mathcal{T}  \int  p(\mathbf{Y}^{\tau}| \mathbf{S}^{\tau},\mathbf{C}^{\tau}, \boldsymbol{\phi}^{\tau} ) p(\boldsymbol{\phi}^{\tau} | \boldsymbol{\theta} ) d \boldsymbol{\phi}^{\tau} \right) d\boldsymbol{\theta}\right] \nonumber \\
\geq & \mathbb{E}_{q(\boldsymbol{\theta};\psi)} \left[ \log \prod_{\tau=1}^\mathcal{T}  \int p(\mathbf{Y}^{\tau}| \mathbf{S}^{\tau},\mathbf{C}^{\tau}, \boldsymbol{\phi}^{\tau} )  p(\boldsymbol{\phi}^{\tau} | \boldsymbol{\theta} ) d \boldsymbol{\phi}^{\tau} \right] \nonumber\\ & - D_{KL}(q(\boldsymbol{\theta};\psi) || p(\boldsymbol{\theta}))\nonumber \\
= & \mathbb{E}_{q(\boldsymbol{\theta};\psi)} \left[\sum_{\tau=1}^\mathcal{T}  \log  \int p(\mathbf{Y}^{\tau}| \mathbf{S}^{\tau},\mathbf{C}^{\tau}, \boldsymbol{\phi}^{\tau} )  p(\boldsymbol{\phi}^{\tau} | \boldsymbol{\theta} ) d \boldsymbol{\phi}^{\tau} \right] \nonumber\\ & - D_{KL}(q(\boldsymbol{\theta};\psi) || p(\boldsymbol{\theta}))\nonumber \\
\geq & \mathbb{E}_{q(\boldsymbol{\theta};\psi)} \left[\sum_{\tau=1}^\mathcal{T} \mathbb{E}_{q(\boldsymbol{\phi}^{\tau};\boldsymbol{\lambda}^{\tau})} \left[ \log p(\mathbf{Y}^{\tau}| \mathbf{S}^{\tau},\mathbf{C}^{\tau}, \boldsymbol{\phi}^{\tau} )   \right] \right. \nonumber\\ & \left. - D_{KL}(q(\boldsymbol{\phi}^{\tau};\boldsymbol{\lambda}^{\tau}) || p(\boldsymbol{\phi}^{\tau} | \boldsymbol{\theta})) \vphantom{\sum_{m=1}^{M^\tau}}
\right] 
- D_{KL}(q(\boldsymbol{\theta};\psi) || p(\boldsymbol{\theta})) \nonumber\\
=& \mathcal{L}\left(\psi, \{\boldsymbol{\lambda}^{\tau}| \tau= 1, \ldots, \mathcal{T}\}\right).
\end{split} 
\end{equation}





	
	\clearpage
	
	
	
	
	\small
	\bibliographystyle{ACM-Reference-Format}


	

	
	

\end{document}